\documentclass[10pt,twocolumn]{article}
\usepackage[textwidth=7in,textheight=8.5in]{geometry}
\usepackage{graphicx}
\usepackage{amsmath} 
\usepackage{amssymb} 
\usepackage{fancyvrb} 
\usepackage{soul}
\usepackage{fancyhdr}
\usepackage{epstopdf}
\usepackage{lineno}
\usepackage{hyperref}
\makeatletter
\renewcommand{\maketitle}{\bgroup
\begin{flushleft}
  \begin{Huge}
  \textbf{\@title}\\
  \end{Huge}
  \vspace{1cm}
  \@author
\end{flushleft}\egroup
}
\makeatother
\title{Symmetry-Scaling Based Complex Network Approach to Explore Exotic Hadronic States in High-Energy Collision.}
\author{%
    \textbf{{\large Susmita Bhaduri}}$^{1}$,\textbf{{\large Anirban Bhaduri}}$^{2}$ \textbf{{\Large Dipak Ghosh}}$^{3}$\\
    $^{1,2,3}$Deepa Ghosh Research Foundation, Kolkata-700031,India \\
    \underline{$^{1}$susmita.sbhaduri@dgfoundation.in}\\
    \underline{$^{2}$bhaduri.anirban@dgfoundation.in}\\
    \underline{$^{3}$dipak.ghosh@dgfoundation.in}
}

\begin{document}
\twocolumn[
  \begin{@twocolumnfalse}
    \maketitle
  \end{@twocolumnfalse}
  ]
\noindent
\begin{abstract}
Conventionally invariant mass or transverse momentum techniques have been used to probe for any formation of some exotic or unusual resonance states in high energy collision. In this work, we have applied symmetry scaling based complex network approach to study exotic resonance/hadronic states utilizing the clustering coefficients and associated scaling parameter extracted with the complex network based technique of Visibility Graph. We have analyzed the data of Pb-Pb collision data sample at $2.76 TeV$ from ALICE Collaboration and analyzed different patterns of symmetry scaling, scale-freeness, correlation and clustering among the produced particles. This is a chaos-based complex network technique where simple parameters like Average Clustering Coefficient and Power of Scale-freeness of Visibility Graph(PSVG) may hint at formation of some exotic or unusual resonance states without using conventional methods. From this experiment we may infer that highest range of Average clustering coefficient, might be the resonance states/clusters from where the hadronic decay might have occurred and few clusters with highest value of this parameter may indicate that those clusters may be the ancestors of some strange particles.
There have also been extensive study of dilepton production since the study of lepton pair generation in Drell-Yan processes is immensely important as these processes enable us to validate the Standard Model-SM predictions for the fundamental particles interaction at the new energy region and also to probe for new physics beyond SM. Hence we have applied the same methodology and extracted the same parameters for p-p collision data at $8$TeV from CMS, to detect possible resonance states eventually generating lepton pairs.
\end{abstract}



%

\section{Introduction}
\label{intro}
Numerous analysis on the fluctuation pattern of pions have been studied from theoretical and phenomenological approaches with a view to know the underlying dynamics of pionisation process and also critical phenomena like QGP production. 
Initially a new concept called intermittency was introduced to study large multiplicity fluctuations with correlation fluctuation which yielded encouraging results. Also, it has been observed that multipion production shows a power-law behavior of the factorial moments with respect to the size of phase-space intervals in decreasing mode. An indication of a self-similar fluctuation was thereby obtained, which in turn indicated the fractal behaviors in statistical and geometrical systems. 
The study of fractal behavior of multipion production from the perspective of intermittent fluctuations using the method of factorial moment, had been an initial area of interest. 
Then, numerous techniques based on the fractal theory had been implemented to analyse the process of multipion emission in terms of Gq moment and Tq moment~\cite{hwa90,paladin1987,Grass1984,hal1986,taka1994}. Then techniques like the Detrended Fluctuation Analysis(DFA), multifractal-DFA (MF-DFA) method~\cite{cpeng1994, kantel2002} have been applied extensively for analyzing nonlinear, non-stationary data series for identifying their long-range correlations.
A number of multifractal analysis of particle production processes has been also reported in the recent times~\cite{Albajar1992,Suleymanov2003,YXZhang2007,Ferreiro2012}. At the same time self-similarity has been studied in different works of particle physics, e.g, in the process of Jet and Top-quark production at TEVATRON and LHC~\cite{Tokarev2015}, strangeness production in pp collisions at RHIC~\cite{Tokarev2016}, proton spin and asymmetry of jet production~\cite{Tokarev20151}, for describing collective phenomena~\cite{Baldina2017} and also in the application of self-similar symmetry model to dark energy~\cite{TomohideSonoda2018}.

A major shift of pradigm ocuured with the latest advances in the field of complex network. Albert and Barab{\'{a}} have studied latest works in this field and examined the analytic tools and models for random graphs, small-world and scale-free networks, in the recent past~\cite{Albert2002,Barabasi2011}. 
Lacasa et al. have introduced Visibility Graph analysis~\cite{laca2008,laca2009} method which has gained importance due its entirely different, rigorous approach to assess fractality. Lacasa et al. have analysed real time series in different scientific fields, using fractional Brownian motion(fBm) and fractional Gaussian noises(fGn) series as a theoretical framework. Lacasa et al. mapped fractional Brownian motion(fBm) and fractional Gaussian noises(fGn) series into a scale-free Visibility Graph having the degree distribution as a function of the Hurst exponent~\cite{laca2009}. This way, they have applied classical method of complex network analysis to unambiguously quantify long-range dependence and fractality of a time series~\cite{laca2009}. 
This method has been used productively for analyzing various biological signals in recent works \cite{Bhaduri2014,Bhaduri20163,nil2016,bhaduriJneuro2016,Bhaduri2017}. 
In view of the above, recently, using Visibility Graph method, we have analysed multiplicity fluctuation in hadron-nucleus and nucleus-nucleus interactions~\cite{Bhaduri20167,Bhaduri20171,Bhaduri20183,Bhaduri20163,Bhaduri20165,Bhaduri20166,Bhaduri20172,Bhaduri20181}.
In our latest work~\cite{Bhaduri2630203,Bhaduri20184}, we have done the scaling analysis of the pseudorapidity data extracted from Pb-Pb VSD masterclass data sample at $2.76 TeV$ per nucleon pair from ALICE Collaboration~\cite{alice} using the method of Visibility Graph from the perspective of complex network and the multifractal methodologies, to probe for phase transition and capture the signature of QGP. 

After initial success in study of pionisation process in high energy collision with chaos-based complex network approach we have ventured to study exotic resonance/hadronic states utilizing the clustering coefficients and associated symmetry-scaling based parameter extracted with the complex network based technique of Visibility Graph. We have framed a new methodology to identify clusters which are ancestors of hadronic decay product in this case. We intend to propose a novel chaos-based complex network technique where simple parameters may hint at formation of some exotic or unusual resonance states without using conventional invariant mass or transverse momentum techniques. 
In this work we have extracted Average clustering coefficients from Pb-Pb VSD masterclass data sample at $2.76 TeV$ per nucleon pair from ALICE Collaboration~\cite{alice} and we have observed that scale-free clusters are being created with substantially high Average clustering coefficients. From this we may infer that all the clusters with higher range of Average clustering coefficient, might be the resonance states from where the hadronic decay might have occurred and few clusters with highest value of this parameter may indicate that those few clusters might have been ancestors of some strange particles.

We also performed similar analysis in another important domain of high energy interaction namely dilepton production since the study of lepton pair generation in Drell-Yan processes is immensely important as these processes enable us to validate the Standard Model-SM predictions for the fundamental particles interaction at the new energy region and also to probe for new physics beyond SM. The Drell-Yan process was the first implementation of parton model ideas beyond deep inelastic scattering. We have used the dimuon data extracted from the primary dataset of RunB of the p-p collision at $8$TeV from CMS collaboration~\cite{cms2017}, for this experiment and attempted to detect possible resonance states eventually generating lepton pairs, using the proposed methodology and the novel parameters representing correlation/clustering and degree of symmetry scaling. 

The rest of the paper is organized as follows. The methods of analysis are described in Section~\ref{ana} - Visibility Graph algorithm, the significance of its scale-freeness property and its Average Clustering Coefficient are presented in Section~\ref{ana}. The data description is there in Section~\ref{data}. The objective of this experiment, the step-by-step details the analysis and the analysis of the test results are given in Section~\ref{method}. The physical significance of the observable parameters and their significance with respect to the resonance states or exotic states, is elaborated and the paper is concluded in Section~\ref{con}.

\section{Method of analysis}
\label{ana}
We would be describing complex network-based method of Visibility Graph algorithm and the significance of its symmetry scaling or scale-freeness property to extract the PSVG(Power of Scale-freeness of Visibility Graph)~\cite{laca2008,laca2009}, and another parameter known as Average Clustering Coefficient of the Visibility Graph in this section. We would be using these parameters for analyzing the fluctuation of pseudorapidity space($\eta$-space) of the event datasets extracted from the experimental data.

\subsection{Visibility Graph Algorithm}~\\
\label{vgalgo}
\begin{figure}[h]
\centerline{
\includegraphics[width=3in]{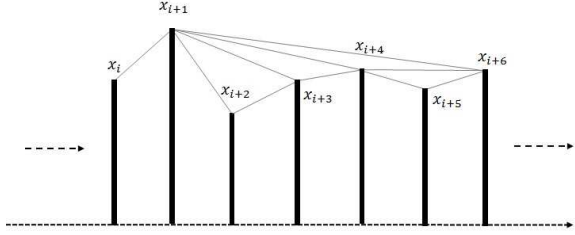}
}
\caption{Visibility Graph for time series X}
\label{visi}
\end{figure}
Visibility Graph algorithm plots time series $X$ to its Visibility Graph. Lets say, the $i^th$ point of time series is $X_{i}$. In this graph two nodes or vertices($X_{m}$ and  $X_{n}$) are meant to be connected by a two-way edge if and only if the equation~\ref{ve} is valid.

\begin{equation}
X_{m+j} < X_{n} + (\frac{n - (m+j)}{n-m})\cdot(X_m - X_n) \nonumber
\label{ve}
\end{equation}
\begin{math}
\mbox{where }
\forall j \in Z^{+} \mbox{ and } j < (n-m)\\
\\
\end{math} 
Fig.~\ref{visi} shows that the nodes $X_{m}$ and  $X_{n}$, where $m=i$ and $n=i+6$, are visible to each other only if the Eq. \ref{ve} is valid. 
It is obvious that two consecutive points of the time series can always see each other and thereby sequential nodes are always connected.

\subsubsection{Network parameters}~\\
\label{nwparm}
The network parameters to be extracted from the scale-free Visibility Graphs formed from each of the experimental data series are elaborated below.
\begin{enumerate}
\item \textbf{Power of Scale-freeness of VG \mbox{-} PSVG :}
\label{psvg}
The degree of a node or vertex in a graph \mbox{-} here Visibility Graph, is the number of links or edges the node has with the rest of the nodes in that graph. The degree distribution $P(k)$ of a network is therefore defined as the fraction of nodes with degree $k$, with respect to the total number of nodes present in the network. So, if there are $n_k$ number of nodes in the network, having degree $k$ and total number of nodes in total in a network is $n$, then we define $P(k) = n_k/n$ for all possible values of $k$. 

As per Lacasa et al.\cite{laca2008,laca2009} and Ahmadlou et al.\cite{ahmad2012}, the degree of scale-freeness of a Visibility Graph corresponds to the amount of fractality and complexity of the time series.
According to the scale-freeness property of Visibility Graph, the degree distribution of its nodes should follow power-law, i,e, $P(k) \sim k^{-\lambda_p}$, where $\lambda_p$ is a constant and it is called the \textbf{Power of the Scale-freeness in Visibility Graph-PSVG}. 
Hence $\lambda_p$ or the PSVG measures the amount or degree of self-similarity, fractality and it's also the estimation of complexity of the time series. As the fractal dimension estimates the amount of self-similarity of a time series, $\lambda_p$ indicates the FD \mbox{-} Fractal Dimension of the signal\cite{laca2008,laca2009,ahmad2012}. 
It has also been observed that there exists an inverse linear relationship between PSVG-$\lambda_p$ and Hurst exponent of the associated time series\cite{laca2009}.

\item \textbf{Average Clustering Coefficient :} 
Clustering coefficient is the estimation of the extent to which nodes of a graph tend to cluster together. Average-clustering-coefficient had been defined by Watts and Strogatz~\cite{Watts1998}, as the comprehensive clustering coefficient of a network, which is calculated as the average local clustering coefficient of all the nodes in the graph. 

A graph/network $G=(V,E)$ is made up of a set of vertices $V$ and another set of edges $E$ between the vertices. An edge $e_{ij} \in E$, links vertices $v_{i}$ and $v_{j}$. The neighborhood, say denoted by $N_{i}$, for any vertex $v_{i} \in V$ is determined as its neighbors which are immediately linked to it, say denoted by, $N_{i}=\{v_j:e_{ij} \in E \vee e_{ji} \in E\}$. Strogatz and Watts have determined the local-clustering-coefficient, say denoted by $C_i$ for the vertex $v_i$, as the number of links existing between the vertices inside its neighborhood, divided by the number of links that could possibly exist between them. For the directed graphs/networks, $e_{ij}$ is different from $e_{ji}$, so for every neighborhood $N_i$ corresponding to each vertex $v_i$, there can exist maximum of $k_i(k_i-1)$ number of links amid the vertices inside $N_i$, where $k_i$ is the number of elements in the set $N_i$, or the number of neighbors of $v_i$. In this case $C_i$ is determined as~$\frac{\left\vert{\{e_{jk}:v_j,v_k \in N_i,e_{jk} \in E\}}\right\vert}{k_i(k_i-1)}$.

However, for undirected graphs/networks $e_{ij}$ is considered as identical to $e_{ji}$. So, as $v_i$ has $k_i$ number of neighbors within $N_i$ and a maximum of $\frac{k_i(k_i-1)}{2}$ number of edges can exist among the vertices inside $N_i$. 
So here $C_i$ is determined as~$\frac{2\left\vert{\{e_{jk}:v_j,v_k \in N_i,e_{jk} \in E\}}\right\vert}{k_i(k_i-1)}$.

Then the Average-clustering-coefficient of a graph/network, say denoted by $\bar{C}$, is estimated by averaging the local clustering coefficients for all the vertices in the graph/network. High value of this parameter suggests high robustness of a graph/network.

\end{enumerate}

\section{Data description}
\label{data}
Two sets of data has been analyzed with the proposed methodology. The steps for the analysis is described in detail in the Section~\ref{method}.

\subsection{Pimesons from Pb-Pb collision data at $2.76$ TeV per nucleon pair from ALICE}
\label{alidata}
In this analysis, using CERN Root software, we have extracted $10$ root dataset files, namely AliVSD\_Masterclass\_$1,2,\ldots,10$, each containing $34$ event datasets summing upto a total of $340$ event datasets from Pb-Pb VSD masterclass data sample at $2.76 TeV$ per nucleon pair from ALICE Collaboration~\cite{alice}. The derived dataset has been downloaded from CERN Open Data Portal (http://opendata.cern.ch/record/1120) which contains the above mentioned $10$ root dataset files. 
Derived datasets as defined by CERN are the datasets which contain data that have been derived from the primary datasets. The data may be reduced in the sense that either only part of the information is kept or only part of the events are selected. In our case the $14$ primary datasets are also available at CERN Open Data Portal in this \href{http://opendata.cern.ch/search?page=1&size=20&type=Dataset&experiment=ALICE&subtype=Collision}{\textit{link}}.

AliVSD\_Masterclass\_$1,2,\ldots,10$ - these root datasets from Pb-Pb VSD masterclass data sample at $2.76 TeV$ per nucleon pair have been processed using the Root software. 
Then from each of the $340$ event datasets, the pseudorapidity space is extracted. Hence in effect we extracted $34$, $\eta$-space from each of the root dataset. Finally, total of $34 \times 10 = 340$ number of $\eta$-spaces are obtained. 
For each of these $340$, $\eta$-space is created by putting $\eta$ values in a sequence. Along the $X$-axis the sequence number is plotted and each $\eta$-value corresponding to the sequence is plotted along the $Y$ axis. This way we have extracted single $\eta$ data series for each of the $340$ event-dataset.

\subsection{Dimuon data from p-p collision at $8$ TeV from CMS collaboration}
\label{cmsdata}
The primary dataset of the p-p collision at $8$TeV from CMS collaboration contains dataset AOD format from RunB of 2012~\cite{cms2017}. From there the luminosity sections in which runs are considered good are extracted and from those runs we have extracted a list of pseudorapidity values for the produced dimuons. For this experiment we have taken a list of pseudorapidity values from the primary dataset containing a number of events. 

\section{Method of analysis and results}
\label{method}
\subsection{Objective of the Experiment}
\label{physics}
Here, we propose a completely different approach based on the study of self-similarity/symmetry-scaling inherent in exotic resonance/hadronic states utilizing the properties of clusters extracted from the self-similar Visbility Graphs generated from the  pseudorapidity parameters of the interactions as described in Section ~\ref{data}. The methods of identifying exotic resonance/hadronic states is based on self-similarity of symmetry-scaling and estimation of clustering coefficients. As mentioned earlier the concept of self-similarity in the context of high energy interaction has been applied in cases like strangeness production in pp collisions~\cite{Tokarev2016}, Top-quark production~\cite{Tokarev2015}, proton spin and asymmetry of jet production~\cite{Tokarev20151} etc. In the similar perspective we attempt to explore the possibility of identifying exotic resonance/hadronic states in p-p collision and Pb-Pb collision in recent LHC experiments. The final state hadrons or muons are analyzed following the procedure described in Section~\ref{steps} to find out the Power of Scale-freeness of VG \mbox{-} PSVG to obtain the amount of self-similarity - or in other words the fractality of the clusters with higher range of Average Clustering Coefficient, where the clusters are formed out of scale-free Visibility Graphs constructed out of the observable(pseudorapidity) of the interactions from LHC experiments~\ref{data}. 

Total $341$ datasets of pseudorapidity values are extracted from two different experiments as specified below.
\begin{itemize}
\item $34$ event datasets are extracted from each of the $10$ root datasets from ALICE collaboration~\cite{alice}, as described in Section~\ref{alidata}. So in total $340$ pseudorapidity-datasets are extracted here.
\item One dataset containing pseudorapidity values from multiple validated runs extracted from the root datasets of RunB of the p-p collision at $8$TeV from CMS collaboration~\cite{cms2017}, as described in Section~\ref{cmsdata}
\end{itemize}

Average Clustering Coefficient and PSVG are extracted from the two sets of data series as per the process described in Section~\ref{ana} and step-by-step method of analysis is elaborated subsequently.
\subsection{Step-by-step Method of Analysis}
\label{steps}

\begin{enumerate}
\item Firstly $341$ Visibility Graphs formed out of $341$ datasets extracted from Pb-Pb data sample at $2.76 TeV$ per nucleon pair from ALICE and p-p collision dimuon dataset of $8$ TeV, as listed above. Visibility Graph is constructed as per the method described in Section~\ref{vgalgo}.
\item Then for each of the $341$ Visibility Graphs, $10$ clusters are extracted. In this experiment the clusters are created based on two factors - the visibility between the nodes of the constructed Visibility Graph and the density of the nodes, i.e. how close the nodes of the Graph are in terms of the distance or number of edges between them. The nodes which are visible to each other as well as close to each other in the Visibility Graph are included in the same cluster. The Figure~\ref{clusters_alice} and~\ref{clusters_cms} show a sample set of $5$ clusters extracted from the dataset of $\eta$-values from one of the Pb-Pb dataset from ALICE Collaboration and p-p dataset from CMS collaboration respectively.

\begin{figure*}[t]
\centerline{\includegraphics[width=6in]{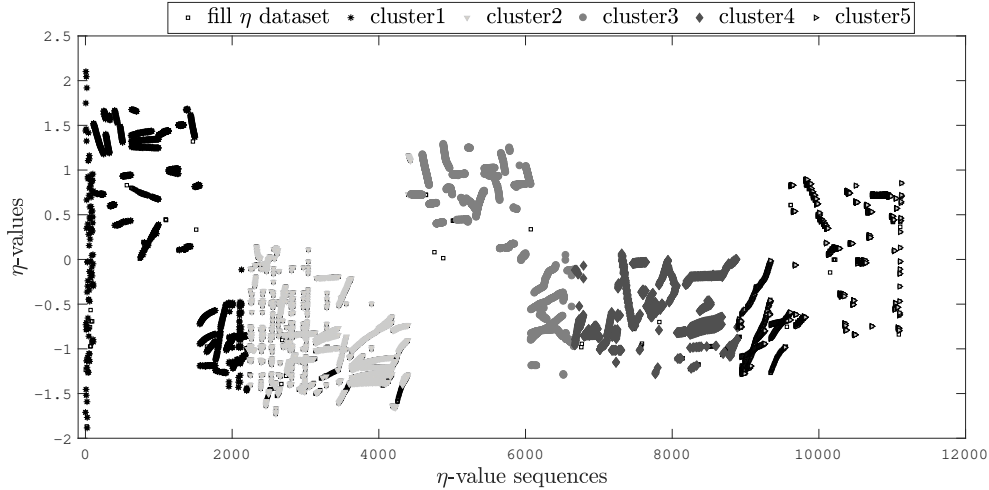}}
\vspace*{8pt}
\caption{$5$ clusters extracted from a Pb-Pb data sample at $2.76 TeV$ per nucleon pair from ALICE Collaboration.\protect\label{clusters_alice}}
\end{figure*}

\begin{figure*}
\centerline{\includegraphics[width=6in]{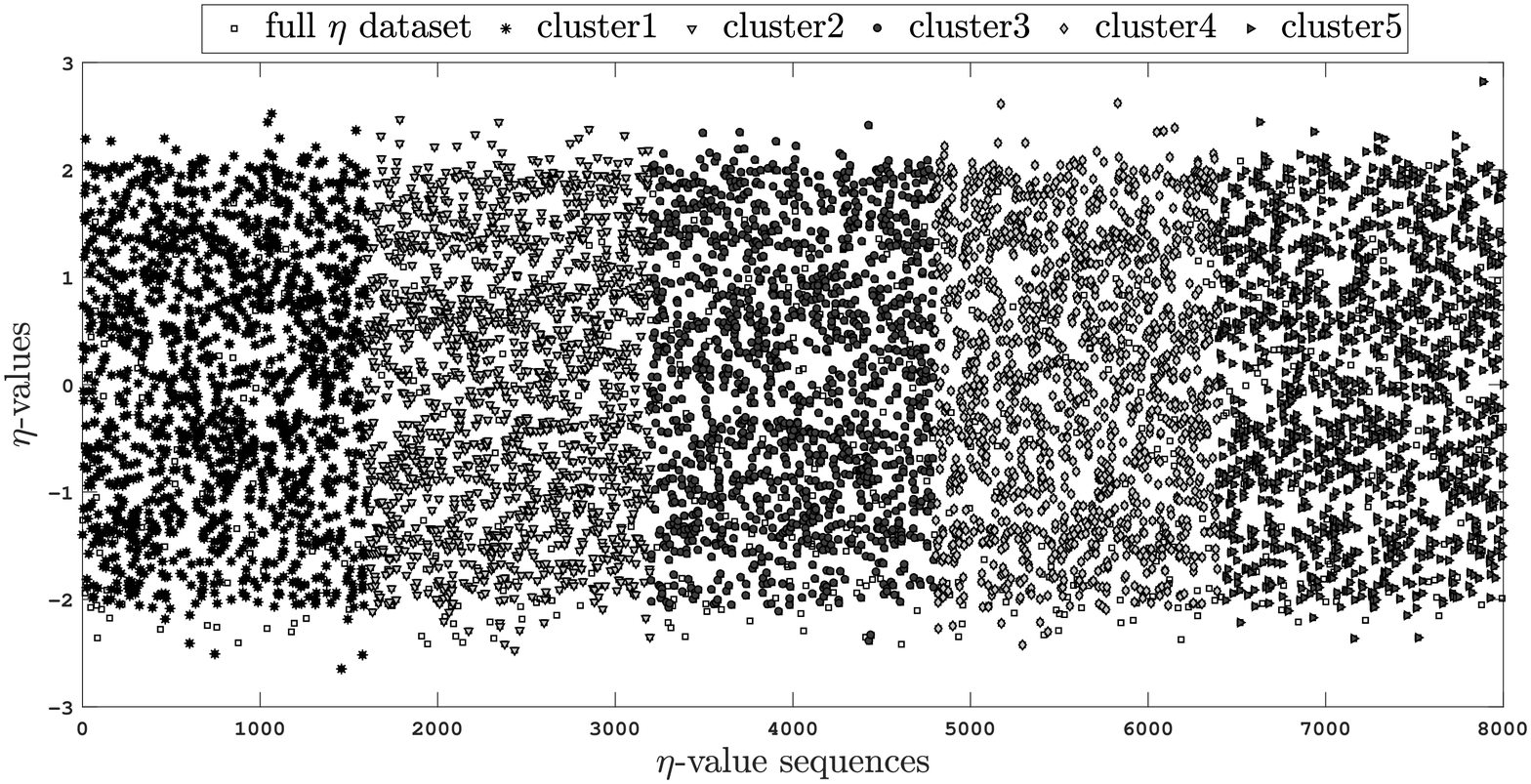}}
\vspace*{8pt}
\caption{$5$ clusters extracted from p-p collision at $8$TeV from CMS collaboration.\protect\label{clusters_cms}}
\end{figure*}

\item Once the clusters are formed, i.e. $10$ clusters from each of the $341$ Visibility Graphs, then for each cluster again scale-free Visibility Graph is constructed following the method elaborated in Section~\ref{vgalgo}. From that graph the parameters of Average Clustering Coefficient and PSVG for each Visibility Graphs is calculated. as per the method prescribed by Watts and Strogatz~\cite{Watts1998} which is described in Section~\ref{nwparm}. 

\item For all the scale-free Visibility graphs constructed from the clusters, the network parameters are calculated and analyzed. 
\begin{enumerate}

\item \textbf{Average Clustering Coefficient:} 
Here the criteria for clustering is the tendency of the nodes in the Visibility Graphs, to be visible to each other. Hence the existence of path or edge between neighbors of a particular node in the Graph, is decided by their visibility to each other. The more visible the neighbor nodes are to each other, the more correlated and clustered they are, for a particular node. This way for each node, the correlation between its neighbor nodes is calculated for a particular Visibility Graph created for a particular cluster and finally Average-clustering-coefficient of the particular scale-free Visibility Graph or that scale-free cluster is measured. 

As already discussed in the Section~\ref{nwparm}, this parameter indicates the probability that whether the neighbor-nodes of a node in a graph, are also neighbors to each other or not. Average Clustering Coefficient for each Visibility Graphs is calculated as per the method prescribed by Watts and Strogatz~\cite{Watts1998}. 

\begin{figure*}[h]
\centerline{\includegraphics[width=5in]{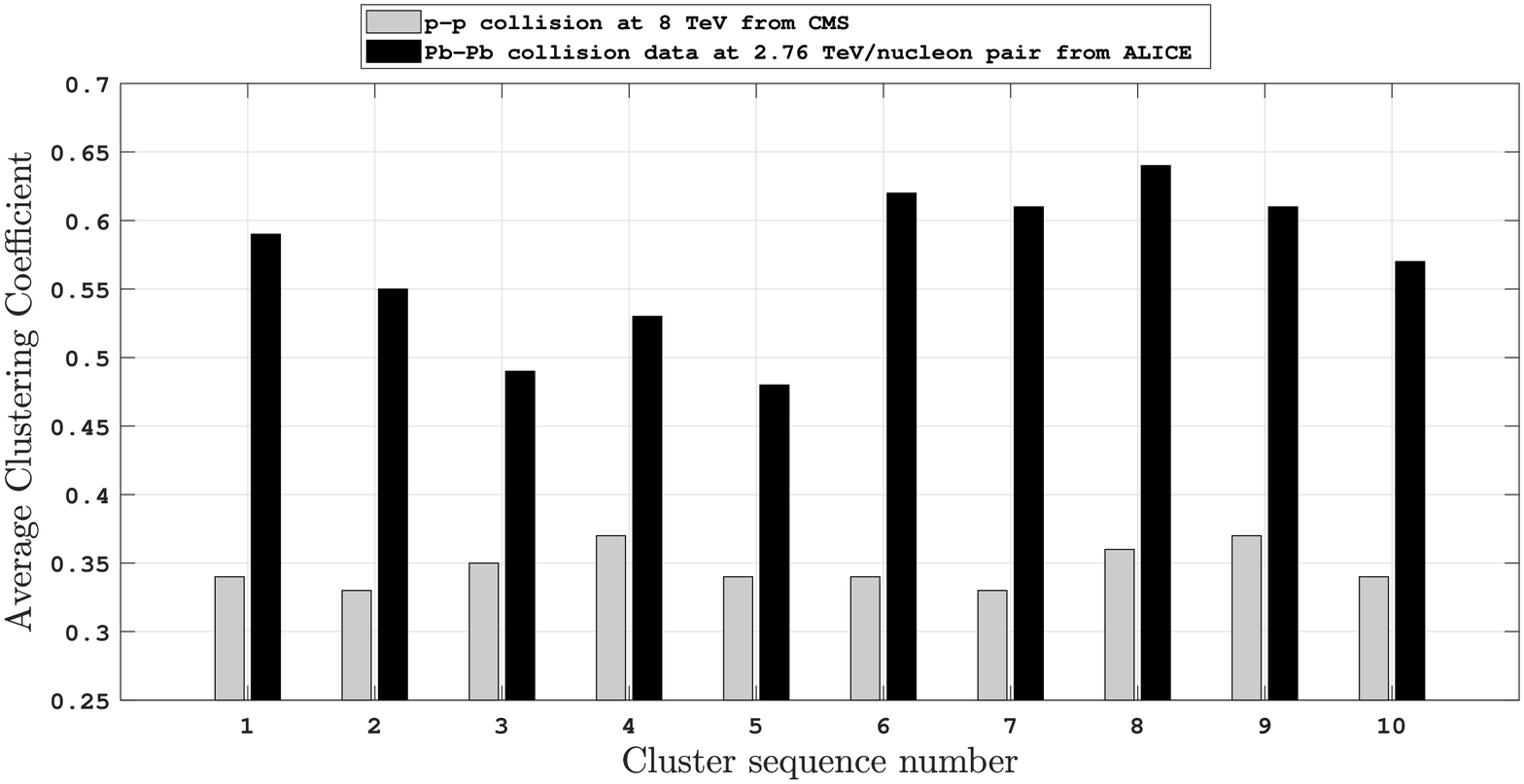}}
\vspace*{8pt}
\caption{Comparison of Average Clustering Coefficients calculated for $10$ clusters extracted from each of a single event dataset of Pb-Pb collision at $2.76 TeV$/nucleon pair from ALICE Collaboration and p-p collision dataset at $8$TeV from CMS collaboration.\protect\label{clustco_comp}}
\end{figure*}

Figure~\ref{clustco_comp} shows the comparison of Average Clustering Coefficients calculated for $10$ scale-free clusters extracted from each of a single event dataset of Pb-Pb collision at $2.76 TeV$/nucleon pair from ALICE Collaboration and p-p collision dataset at $8$TeV from CMS collaboration. Evidently, the clusters of Pb-Pb collision event are substantially more clustered and correlated than those of p-p collision data. Moreover it is evident from the Figure~\ref{clustco_comp}, the maximum Average Clustering Coefficient among the $10$ clusters extracted from p-p collision-dataset at $8$TeV from CMS collaboration is $0.37$ of the cluster number $4$ and $9$. Similarly, the same parameter obtains its maximum value among the $10$ clusters extracted from a Pb-Pb collision-event at $2.76 TeV$/nucleon pair from ALICE Collaboration is $0.64$ of the cluster number $8$.

\item \textbf{Power of Scale-freeness of VG \mbox{-} PSVG :}Values of Power of Scale-freeness of Visibility Graph \mbox{-} PSVG is extracted as per the method described in Section~\ref{psvg}.
For each of the $10$ scale-free Visibility Graphs formed from the clusters extracted from the Visibility Graphs formed out of each of the $341$ event-datasets from ALice and CMS experiments, the $k$ vs $P(k)$ dataset is calculated as per the method described in Section~\ref{psvg}. 

$k$ vs $P(k)$ plot for a cluster dataset extracted from p-p collision-event at $8$TeV from CMS collaboration, is shown in Figure~\ref{power_cms}-(a) and similar plot for Pb-Pb collision-event at $2.76 TeV$/nucleon pair from ALICE collaboration is shown in Figure~\ref{power_ali}-(a). The power-law index has been obtained by power-law fitting for the $k$ vs $P(k)$ datasets as per the method by Clauset et al.~\cite{Clauset2009}. The power-law relationship can be confirmed from the corresponding value of $R^2 = 0.95$ and $0.90$ respectively.

\begin{figure*}[t]
\centerline{
\includegraphics[width=3.5in]{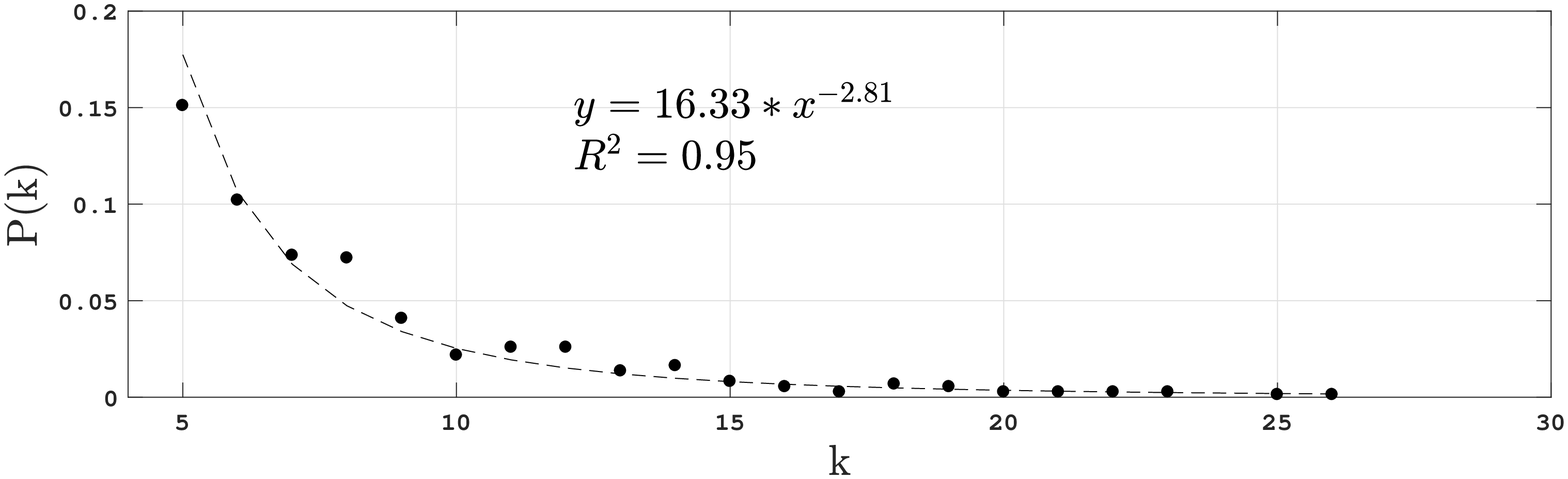}
\includegraphics[width=3.5in]{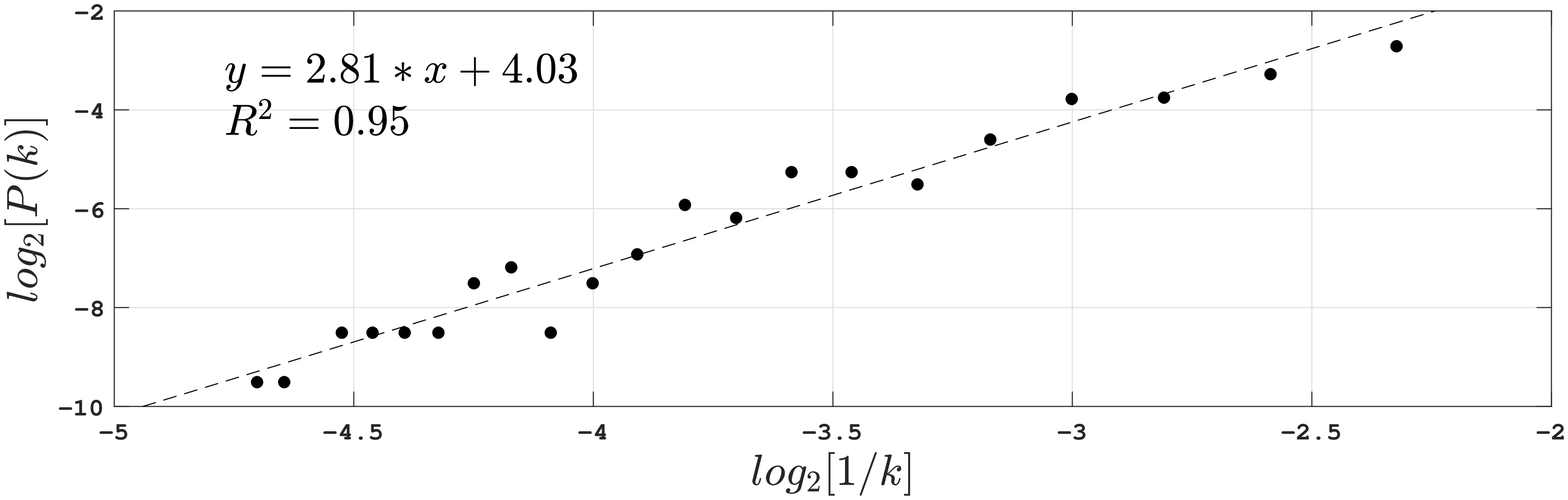}
}
\centerline{(a) \hspace*{6cm} (b)}
\caption{(a) $k$ vs $P(k)$ plot for a cluster dataset extracted from p-p collision-dataset at $8$TeV from CMS collaboration (b) $log_{2}[1/k]$ vs $log_{2}[P(k)]$ plot for the same dataset}
\label{power_cms}
\end{figure*}

\begin{figure*}[h]
\centerline{
\includegraphics[width=3.5in]{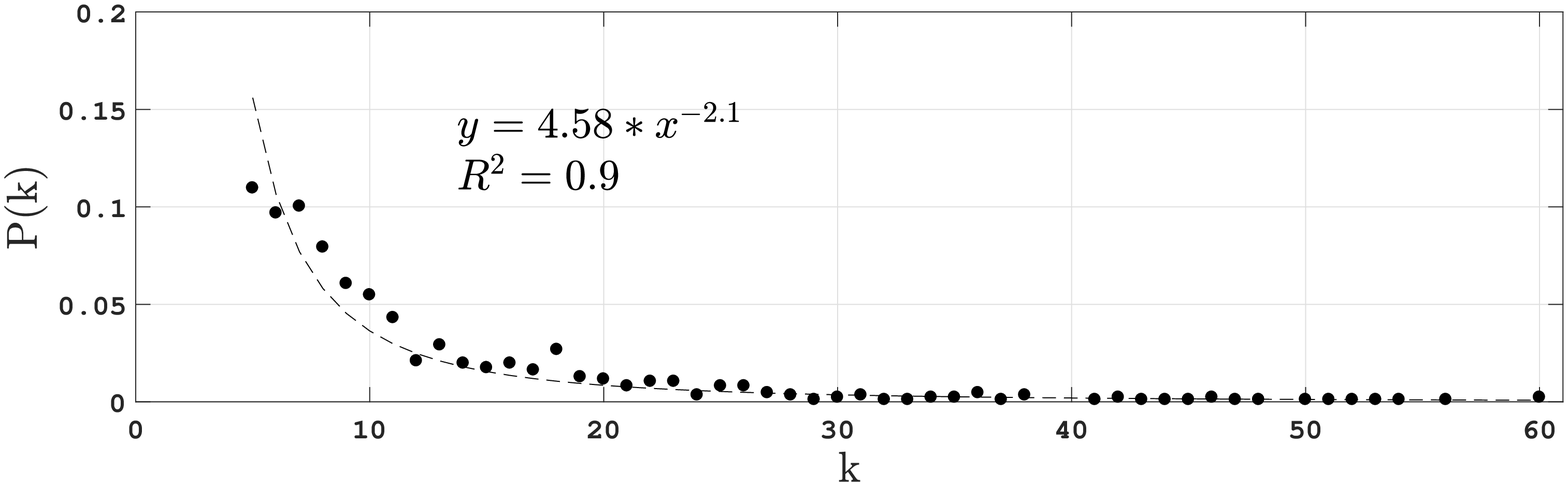}
\includegraphics[width=3.5in]{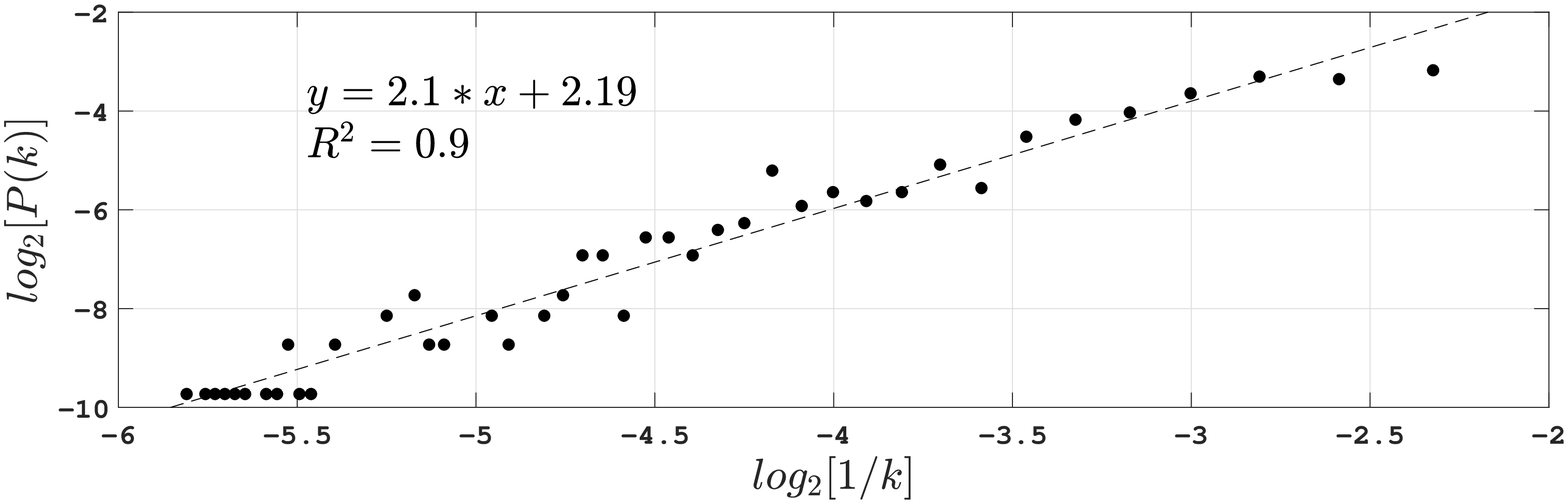}
}
\centerline{(a) \hspace*{6cm} (b)}
\caption{(a) $k$ vs $P(k)$ plot for a cluster dataset extracted from Pb-Pb collision-dataset at $2.76 TeV$/nucleon pair from ALICE collaboration (b) $log_{2}[1/k]$ vs $log_{2}[P(k)]$ plot for the same dataset}
\label{power_ali}
\end{figure*}

The Power of Scale-freeness in Visibility Graph(PSVG), is calculated from the slope of $log_{2}[1/k]$ versus $log_{2}[P(k)]$ for all the Visibility Graphs constructed. Figure~\ref{power_cms}-(b) and~\ref{power_ali}-(b) show the value of the PSVG for the same p-p collision-dataset at $8$TeV from CMS collaboration and Pb-Pb collision-event at $2.76 TeV$/nucleon pair from ALICE collaboration respectively. The high values of $R^2 = 0.95$ and $0.90$ respectively confirm the goodness of straight line fitting. It should be noted that PSVG corresponds to the amount of complexity and fractality of the data series and in turn indicates the fractal dimension of the data series~\cite{laca2008,laca2009,ahmad2012}. It has been observed that, in general, all the scale-free clusters created out of both the LHC events have a moderate to higher degree of scale-freeness.

\end{enumerate}

\end{enumerate}

\subsection{Consolidation of Results from the Analysis}
\label{cons}

\subsubsection{Analysis results of ALICE data}
\label{conali}
For each of the $34$ scale-free Visibility graphs coonstructed from $34$ event datasets corresponding to each of the AliVSD\_Masterclass\_$1,2,\ldots,10$ datasets from Pb-Pb data sample at $2.76 TeV$ per nucleon pair from ALICE collaboration~\cite{alice}, $10$ clusters are extracted and for each cluster again scale-free Visibility Graphs are created as per the steps described in Section~\ref{steps}. Then Average Clustering Coefficient and Power of Scale-freeness of VG \mbox{-} PSVG is calculated for each cluster again as per the process described in the Section~\ref{steps}. After that the results are consolidated as per the following steps.

\begin{enumerate}
\item For each of the AliVSD\_Masterclass\_$1,2,\ldots,10$ datasets, $34*10$ or $340$ scale-free clusters and then same number of Average Clustering Coefficient parameter is calculated. 
\item Then a frequency histogram is generated to consolidate the most clustered or correlated clusters, i.e. clusters having highest range of Average Clustering Coefficient. In essence, for each of the AliVSD\_Masterclass\_$1,2,\ldots,10$ datasets, single  frequency histogram is generated. The Figure~\ref{hist_ali1} shows a frequency histogram corresponding to a sample AliVSD\_Masterclass dataset, containing the Average Clustering Coefficient coefficient values for all the scale-free clusters corresponding to all $34$ event Visibility Graphs for the Masterclass dataset. It is evident from the histogram that the peak bin contains the $116$ clusters with highest range of Average Clustering Coefficient.

\begin{figure*}[h]
\centerline{\includegraphics[width=5in]{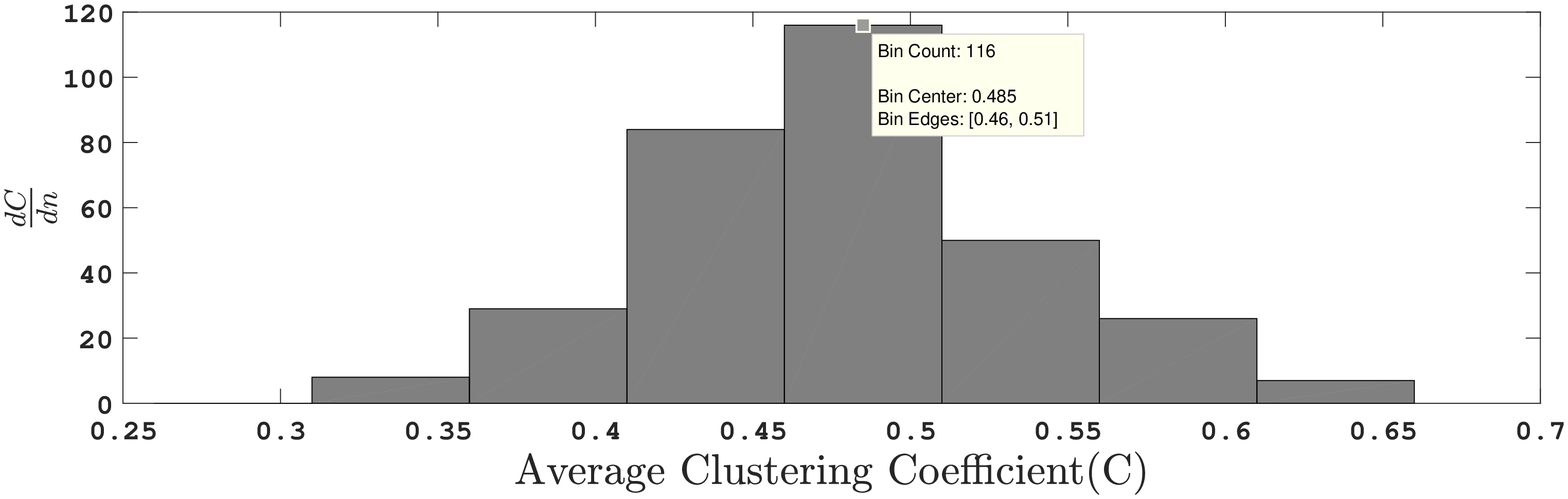}}
\vspace*{8pt}
\caption{Frequency histogram for a sample AliVSD\_Masterclass dataset, containing the Average Clustering Coefficient coefficient values for all the scale-free clusters corresponding to all $34$ event-Visibility Graphs from the data of Pb-Pb collision at $2.76 TeV$/nucleon pair from ALICE Collaboration.\protect\label{hist_ali1}}
\end{figure*}

\item Then the clusters which belong to the peak bin of the histogram due to their highest range of Average Clustering Coefficient values are taken and another frequency histogram is generated from the PSVG values of those clusters.
In this example of AliVSD\_Masterclass dataset, the $116$ scale-free clusters having highest range of Average Clustering Coefficient are taken and another frequency histogram is generated from the PSVG values of those $116$ scale-free clusters. Figure~\ref{histpsvg_ali1} shows the frequency histogram of these $116$ PSVG values. It must be noted that the average PSVG for the $41$ clusters contained the peak bin of the histogram created out of $116$ PSVG values, is around $2.00$.

\begin{figure*}[h]
\centerline{\includegraphics[width=5in]{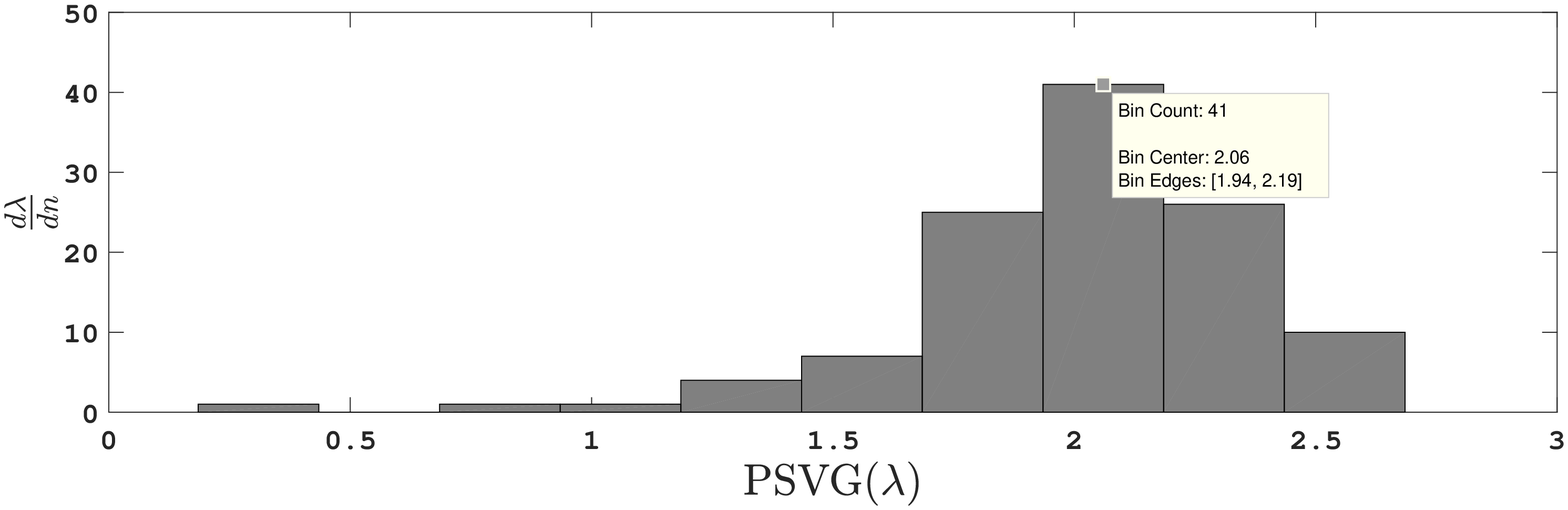}}
\vspace*{8pt}
\caption{Frequency histogram of $116$ PSVG values calculated for $116$ scale-free clusters from the peak bin of the histogram shown in Figure~\ref{hist_ali1} generated for the same sample AliVSD\_Masterclass dataset from the data of Pb-Pb collision at $2.76 TeV$/nucleon pair from ALICE Collaboration.\protect\label{histpsvg_ali1}}
\end{figure*}

\item Step number $1-3$ is repeated for each of the AliVSD\_Masterclass\_$1,2,\ldots,10$ datasets. From the analysis of frequency histograms calculated for the values of Average Clustering Coefficients and PSVG, it has been found that the range of most of the PSVG values calculated for the scale-free clusters is around $1.5-2.3$ though they are highly clustered with a highest range of Average Clustering Coefficient values which is around $0.45-0.60$ for the clusters forming the peak bin of the Average Clustering Coefficient histogram.
\end{enumerate}

\subsubsection{Analysis results of CMS data}
\label{concms}
For the scale-free Visibility Graph extracted from the dimuon event-dataset of pseudorapidity values of CMS data, $10$ clusters are extracted and for each cluster again scale-free Visibility Graphs are created as per the steps described in Section~\ref{steps}. Then Average Clustering Coefficient and Power of Scale-freeness of VG \mbox{-} PSVG is calculated for each cluster again as per the process described in the Section~\ref{steps}. 
Here it is found that the range of PSVG values of the clusters is $2.5-3.0$, though the range of Average Clustering Coefficient is around $0.30-0.35$.

\subsubsection{Results}
\label{inf}
\begin{enumerate}
\item This comparison and analysis done in Section~\ref{conali} and~\ref{concms} show that all the clusters of p-p collision dimuon dataset of $8$ TeV from CMS collaboration, have higher degree of scale-freeness than the ones analyzed for Pb-Pb data sample at $2.76 TeV$ per nucleon pair from ALICE collaboration in the Section~\ref{conali}, though range of Average Clustering Coefficient is substantially low for the CMS data than that of ALICE collaboration. 

Table~\ref{inf_comp} shows the ranges of Average Clustering Coefficient and the PSVG values corresponding to the peak bin of the histogram formed for Average Clustering Coefficient for all the clusters extracted from scale-free Visibility graphs for the collision event data for p-p collision dimuon dataset of $8$ TeV from CMS collaboration and Pb-Pb data sample at $2.76 TeV$ per nucleon pair from ALICE collaboration respectively.

\begin{table*}[h]
\caption{Ranges of Average Clustering Coefficient and the PSVG values corresponding to the peak bin of the histogram formed for Average Clustering Coefficient for all the clusters extracted from scale-free Visibility graphs for the collision event data for p-p collision dimuon dataset of $8$ TeV from CMS collaboration and Pb-Pb data sample at $2.76 TeV$ per nucleon pair from ALICE collaboration respectively.} 
\label{inf_comp}
\begin{tabular}{|c|c|c|}
\hline
Interaction&Average Clustering Coefficient&PSVG values\\
\hline
p-p collision dimuon dataset of $8$ TeV from CMS&$0.30-0.35$&$2.5-3.0$\\
Pb-Pb data sample at $2.76 TeV$/nucleon pair from ALICE&$0.45-0.60$&$1.5-2.3$\\
\hline
\end{tabular}
\end{table*}

\item Higher range of PSVG indicates higher degree of symmetry scaling which is occurring in most of the scale-free clusters extracted for both the interactions from CMS and ALICE collaboration. However for Pb-Pb collision at $2.76 TeV$ per nucleon pair from ALICE collaboration, the hadrons are getting highly correlated or clustered but their symmetry scaling or the scale-freeness is substantially decreasing and sometimes collapsing as evident from poor value of their goodness of fit.

\item Apart from lessening of symmetry scaling, we have observed here that these clusters are becoming highly correlated( as evident from their Average Clustering Coefficient values) for heavy-ion collision of Pb-Pb at a higher energy of $2.76 TeV$/nucleon pair from ALICE. We can infer here that these clusters may be the resonance states from which hadronic decay has occurred.

\item Moreover, for some events of Pb-Pb data sample at $2.76 TeV$/nucleon pair from ALICE, it has been observed that some clusters of some events have unusually higher value of Average clustering coefficient, than specified in Table~\ref{inf_comp}. An example of such one event with its list of clusters and their corresponding Average clustering coefficient is shown in the Figure~\ref{exotic}. Here the $24$th cluster has Average clustering coefficient around $0.70$. This cluster may be indicative of being ancestor of exotic resonance state.

\begin{figure*}
\centerline{\includegraphics[width=5in]{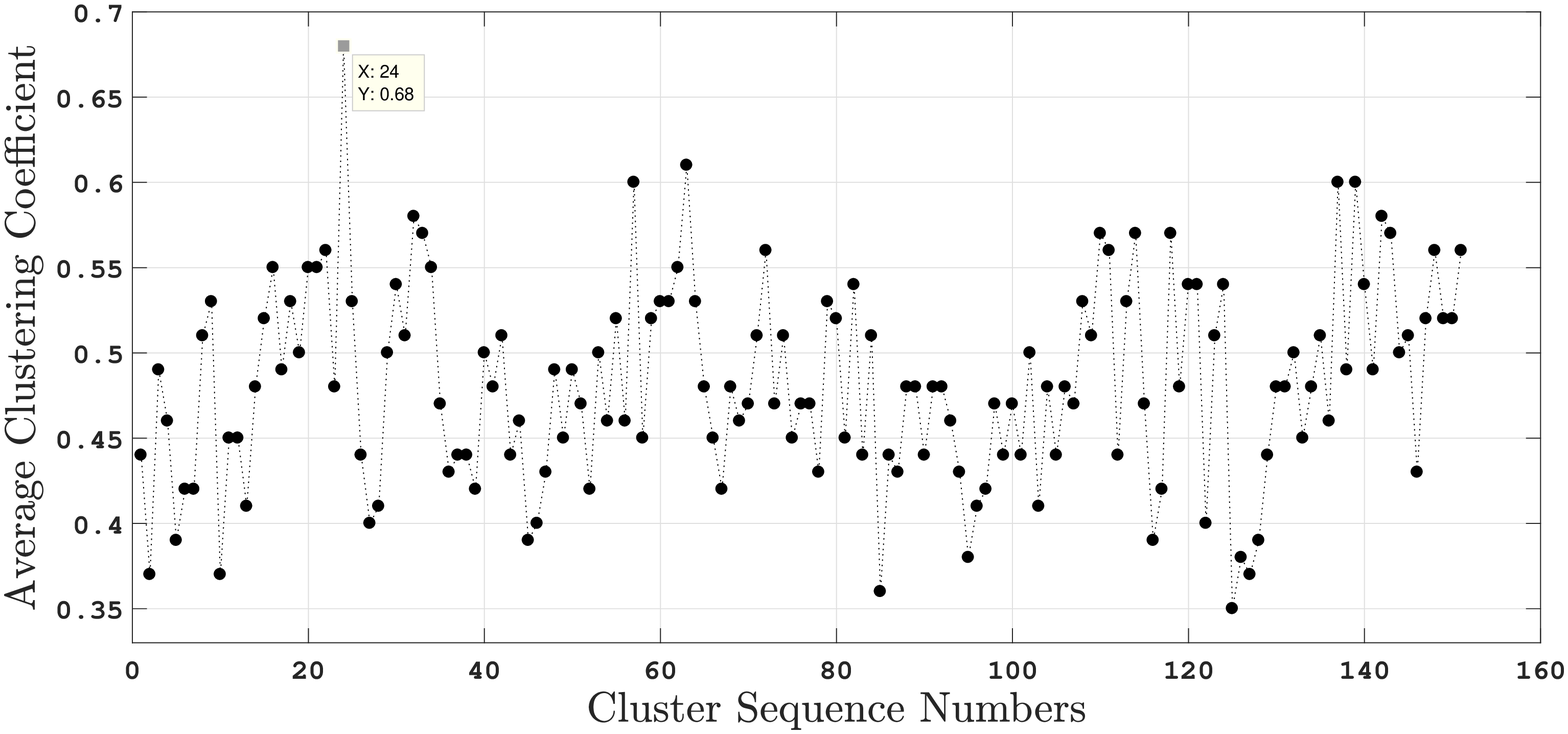}}
\vspace*{8pt}
\caption{Trend of clusters and their corresponding Average clustering coefficient for one event data of Pb-Pb collision at $2.76 TeV$/nucleon pair from ALICE Collaboration.\protect\label{exotic}}
\end{figure*}

\item The same inference may be applicable for p-p collision dimuon dataset of $8$ TeV from CMS and eventually these simple parameters may hint at formation of some resonance states leading to lepton pair generation process without using conventional invariant mass or transverse momentum techniques. 
\end{enumerate}

\section{Conclusion} 
\label{con}
Conventionally invariant mass or transverse momentum techniques have been used to probe for any formation of some exotic or unusual resonance states in high energy collision. In this work, we have applied symmetry scaling based complex network approach to study exotic resonance/hadronic states utilizing the clustering coefficients and associated scaling parameter extracted with the complex network based technique of Visibility Graph. We have analyzed both the pseudorapidity data of Pb-Pb collision data sample at $2.76 TeV$ from ALICE Collaboration using the proposed technique. 
The findings are summarized below.
\begin{itemize}
\item High degree of symmetry scaling has been observed for the scale-free clusters extracted again from the scale-free event datasets from both kind of interactions. Complex network based Visibility Graph method has been applied for this purpose.
\item Substantially higher range of symmetry scaling indicated by the values of the PSVG for the clusters of p-p collision data at $8$TeV from CMS collaboration, though those clusters are substantially lesser correlated or clustered than those of Pb-Pb collision data sample at $2.76 TeV$ from ALICE Collaboration. This may be due to the collapse of symmetry scaling, indicated the poor value of goodness of fit values, for heavy ion collision at much higher energy as found in Alice data.
\item The lesser scale-free clusters are found to be highly correlated or clustered. The higher range of clustering co-efficient might indicate the intermediate resonance states from which hadronic decay may occur in heavy ion high energy interaction. And the very few clusters having even higher Average Clustering Coefficient may be indicative of being ancestor of exotic resonance state.
\item In the similar manner, we may also infer that for high energy interactions like that of p-p collision data at $8$TeV from CMS collaboration, the highly scale-free clusters with highest range of Average Clustering Coefficient may be the clusters of resonance state, and finally after decaying innumerable lepton pairs have been produced from those clusters as they occur in Drell-Yan process conventionally justified with the peaks observed in the invariant mass spectrum of the generated lepton pairs.

\end{itemize}
This summary this symmetry and chaos-based complex network technique of Visibility Graph and the extracted new parameters like Average Clustering Coefficient and Power of Scale-freeness of Visibility Graph may hint at formation of some exotic or unusual resonance states without using conventional methods. Higher range of Average clustering coefficient, might be the resonance states from where the hadronic decay might have occurred and few clusters with highest value of this parameter may indicate that those few clusters might have been ancestors of some strange particles in case of heavy ion high energy interactions.

\section{Acknowledgments} 
\label{ack}
We thank the \textbf{Department of Higher Education, Govt. of West Bengal, India} for logistics support of computational analysis.

\end{document}